\documentclass[a4paper,dvips]{amsart}
\usepackage[dvipdfm]{hyperref}
\usepackage{xcolor}
\usepackage{graphicx}% Include figure files
\usepackage{epsfig}
\definecolor{dark-red}{rgb}{0.4,0.15,0.15}
\definecolor{dark-blue}{rgb}{0.15,0.15,0.4}
\definecolor{medium-blue}{rgb}{0,0,0.5}
\hypersetup{
    colorlinks, linkcolor={dark-red},
    citecolor={dark-blue}, urlcolor={medium-blue}
}
\usepackage{amsmath, amssymb, graphics, setspace}
\usepackage[foot]{amsaddr}	% Part of the standard distribution

\title[Nonlocality vs. UQC in Ising Anyon TQC]{Nonlocality as a Benchmark for Universal Quantum Computation in Ising Anyon Topological Quantum Computers}
\author{Mark Howard$^1$}
\author{Jiri Vala$^{1,2}$}
\address{$^1$ Department of Mathematical Physics, National University of Ireland, Maynooth, Ireland}
\address{$^2$ Dublin Institute for Advanced Studies, School of Theoretical Physics, 10 Burlington Road, Dublin, Ireland}
\email{mark.howard@nuim.ie,Jiri.Vala@nuim.ie}

\DeclareMathOperator{\Tr}{Tr}
\def\ket #1{\vert #1\rangle}
\def\bra #1{\langle #1\vert}
\newcommand{\ketbra}[2]{\ensuremath{\ket{#1}\!\bra{#2}}}

\newcommand{\Jam}{Jamio\l kowski }
\newcommand{\vre}{\ensuremath{\varrho_{\mathcal{E}}}\ }
\newtheorem{theorem}{Theorem}

\begin{document}
\maketitle
\section*{Abstract}
An obstacle affecting any proposal for a topological quantum computer based on Ising anyons is that quasiparticle braiding can only implement a finite (non-universal) set of quantum operations. The computational power of this restricted set of operations (often called stabilizer operations) has been studied in quantum information theory, and it is known that no quantum-computational advantage can be obtained without the help of an additional non-stabilizer operation. Similarly, a bipartite two-qubit system based on Ising anyons cannot exhibit non-locality (in the sense of violating a Bell inequality) when only topologically protected stabilizer operations are performed. To produce correlations that cannot be described by a local hidden variable model again requires the use of a non-stabilizer operation. Using geometric techniques, we relate the sets of operations that enable universal quantum computing (UQC) with those that enable violation of a Bell inequality. Motivated by the fact that non-stabilizer operations are expected to be highly imperfect, our aim is to provide a benchmark for identifying UQC-enabling operations that is both experimentally practical and conceptually simple. We show that any (noisy) single-qubit non-stabilizer operation that, together with perfect stabilizer operations, enables violation of the simplest two-qubit Bell inequality can also be used to enable UQC. This benchmarking requires finding the expectation values of two distinct Pauli measurements on each qubit of a bipartite system.

\section{Introduction and Definitions}

A topological quantum computer (TQC) would allow quantum information to be stored in quasiparticles and manipulated by quasiparticle braiding, in a way that is inherently robust against local perturbations (noise). Many of the most promising candidates for experimental implementation of a useful TQC are those which support Ising anyons, e.g., fractional quantum Hall effect \cite{Moore1991,Bonderson2006}, Majorana wires \cite{Alicea2011}, $p+ip$ superconductors \cite{Ivanov2001} and the Kitaev honeycomb model (in the presence of a magnetic field) \cite{Kitaev2006a}. While there are large differences in the underlying physics of all these systems, the non-Abelian braiding statistics are such that they all possess the same computational power (i.e., they enable the same types of quantum gates and measurements when viewed as a quantum information processing device).

The model of computation (sometimes known as the Clifford computer model \cite{Plenio2010}) that we assume in this work, and the one that is relevant to Ising anyon TQCs, is one in which $(i)$ all single-qubit Clifford gates (discussed in Section \ref{sec:Clifford Operations}), $(ii)$ measurements in the computational basis, and $(iii)$ the controlled-NOT gate, are all implemented in a topologically protected (i.e., effectively perfect) manner. Collectively, we refer to this set of gates and measurements as stabilizer operations. Some proposals for TQCs based on Ising anyons use ``$\ket{a_8}$-distillation'' to achieve the controlled-NOT \cite{Bravyi2006}, while others achieve it using quasiparticle braiding \cite{Georgiev:2006} or non-demolitional measurements of the collective charge of four anyons \cite{Bonderson201x}. It is the latter, topologically protected, implementations of the controlled-NOT that we have in mind here.

A result in quantum information theory -- the Gottesman-Knill theorem (see e.g.~\cite{NielsenChuang:2000} for a discussion) -- says that the set of stabilizer operations is insufficient for achieving universal quantum computation. If a (noiseless) single-qubit unitary gate from outside the Clifford group was implementable then we would immediately have full UQC (this is the case for TQC proposals based on Fibonacci anyons). In fact, for an Ising anyon TQC, attempting to implement any non-stabilizer operation necessitates using non-topological operations which are expected to be highly noisy. For specific target gates and noise models, one can calculate the threshold noise rate \cite{Buhrmanetal:2006,VirmaniHuelgaPlenio:2005,WvDMH:2009} before the power to provide UQC is lost (see e.g., Section \ref{sec:Example: Phase Gate subject to Dephasing} for a relevant example). Rather than consider a number of different target gates (unitaries) and noise models, we examine an overall quantum operation $\mathcal{E}$, which we intend to be non-stabilizer but which may be subject to some unknown evolution, and give an operational benchmark on the utility of $\mathcal{E}$ for UQC. Proving that an operation, $\mathcal{E}$, enables UQC, when used alongside perfect stabilizer operations, reduces to the question of whether $\mathcal{E}$ can be used to produce ancillas that are suitable for magic state distillation \cite{Knill05,BravyiKitaev:2005} (MSD). MSD is a subroutine used in many proposals for fault-tolerant UQC, whereby the requisite non-Clifford gate, $U$, is implemented with the assistance of a non-stabilizer pure qubit state $\ket{\psi_{U}}$. Crucially, many impure (noisy) copies of $\ket{\psi_{U}}$ can be used to create a smaller number of purer $\ket{\psi_{U}}$ \emph{using only stabilizer operations}, and this process can be iterated a number of times if necessary. However, if the ancillas $\ket{\psi_{U}}$ are too impure then it becomes futile to attempt an approximation of $U$, regardless of the number of ancillas and techniques used. A summary of the current methods and limits of MSD can be found in \cite{ReichardtMagic09,CampbellBrowne:2009,CampbellBrowne:2010}.

The benchmark (for identifying $\mathcal{E}$ that enable UQC) that we have just described will involve the violation of a Bell inequality by a two-qubit quantum state, wherein the underlying qubit encoding is provided for by non-abelian Ising anyons. The topological phase supporting Ising anyons is inherently nonlocal, but for the purposes of quantum information processing we are more interested in displaying the nonlocality of encoded qubits. Investigations of nonlocality in non-abelian anyons have been carried out in \cite{Brennen2009,Deng2010} from a different perspective. The results here are motivated mainly by Ising anyon proposals for a TQC, but we note that any system in which magic state distillation is used to achieve fault-tolerant quantum computation assumes that stabilizer operations are effectively perfect (albeit at an encoded level within a CSS error-correcting code, for example). Consequently, the results presented here can be applied to any such system.

Given the restriction on topologically implementable gates, and hence on measurement directions, one quickly sees (see e.g., Section \ref{sec:Nonlocal correlations with restricted operations} later) that this setup precludes the possibility of performing measurements that lead to violation of a Bell inequality. In other words, all measurement outcomes can be described by a local hidden variable (LHV) theory, when we restrict the experimenter to using only topologically protected gates. Previously, we have motivated examining the set of operations $\mathcal{E}$ that can provide UQC. Similarly, we can examine the set of operations $\mathcal{E}$ that can enable violation of a Bell inequality, when used in conjunction with topological operations. Here we relate the two sets of operations with each other, and show that, under certain restrictions, one is strictly included within the other. More generally, we show that any operation that can be used to violate a so-called CHSH Bell inequality can also be used to provide UQC. %If such a nonlocality experiment were performed in an Ising anyon TQC, then one could simultaneously provide

Note that violation of Bell inequalities has been related to quantum computational power in other contexts \cite{Ratanje2011,Hoban2011}, although these works were more focused on generalized entanglement and measurement-based quantum computation respectively. While an interesting topic in its own right, our results do not claim to relate nonlocality and UQC in a broad sense. Indeed in the Ising anyon context, if we are allowed to construct tripartite states (e.g.,  the GHZ state $(\ket{000}+\ket{111})/\sqrt{2}$) then topologically protected operations \emph{are} sufficient to exhibit tripartite nonlocality (e.g. Mermin's experiment \cite{Mermin1990}), whilst still not enabling UQC.

\subsection{Clifford Operations}\label{sec:Clifford Operations}
The single qubit Clifford group has $24$ distinct elements and contains the Pauli group (of order $4$) as a subgroup (strictly speaking we are discussing the Clifford and Pauli groups modulo their center, which amounts to ignoring a global phase in the matrix representation of these elements). Canonical, non-Pauli, elements of the Clifford group are
\begin{align}
H=\frac{1}{\sqrt{2}}\left(
                      \begin{array}{cc}
                        1 & 1 \\
                        1 & -1 \\
                      \end{array}
                    \right),\quad S=\left(
                                 \begin{array}{cc}
                                   1 & 0 \\
                                   0 & i \\
                                 \end{array}
                               \right)
\end{align}\label{eqn:HandS}
and these two operations are actually sufficient to generate the whole group. The characteristic property of Clifford gates, $C$, is that they map Pauli operators to Pauli operators under conjugation, i.e.,
\begin{align*}
C \sigma_j C^\dag \mapsto \pm \sigma_k \quad j,k \in \{x,y,z\}
\end{align*}
When visualized as operations on the Bloch sphere, the Clifford group can be identified with the $24$ possible transformations that can be constructed by combining consecutive $90^\circ$ rotations about the $x$-,$y$- and $z$- axes.

\subsection{Quantum Operations}
To describe the unknown evolution of an Ising anyon qubit, whilst we attempt to implement the requisite non-stabilizer operation, we use the quantum operations formalism (as described in e.g. \cite{NielsenChuang:2000}), where $\mathcal{E}$ is a superoperator that maps input density matrices to output density matrices:
\begin{align*}
\mathcal{E}(\rho_{in})=\rho_{out}.
\end{align*}
A well known tool in quantum information, the \Jam isomorphism, tells us that $\mathcal{E}$ is completely characterized by the output state, \vre, of a process whereby $\mathcal{E}$ is applied to one half of a maximally entangled pair
\begin{align}
\vre=\left(\mathcal{I}\otimes\mathcal{E}\right)\left[\ketbra{\Phi}{\Phi}\right] \quad \text{ where } \Phi=\tfrac{\ket{00}+\ket{11}}{\sqrt{2}} \label{eqn:JamIso}
\end{align}
Because \vre is our preferred method for representing general quantum operations, we will sometimes use $\mathcal{E}$ and \vre interchangeably in latter sections of the paper.

Another convenient way of expressing a quantum operation is in terms of its so-called Kraus operators, $\{E_i\}$, via
\begin{align*}
\mathcal{E}(\rho_{in})=\rho_{out}=\sum_{i}E_i \rho_{in} E_i^\dag \quad (\text{where  } \sum_i E_i^\dag E_i=\mathbb{I})
\end{align*}
It takes $12$ real parameters to completely characterize an arbitrary completely-positive trace-preserving operation. A $9$-parameter subset of the set of all possible operations is given by those that preserve the identity (these are often called unital channels) i.e., those for which $\mathcal{E}(\mathbb{I})= \mathbb{I}$. For unital operations, a Kraus-like description is possible, except the operation is now a probabilistic mixture of unitaries (where each unitary $U_k$ is applied with probability $p_k$)
\begin{align*}
\mathcal{E}(\rho_{in})=\rho_{out}=\sum_{i}p_i U_i \rho_{in} U_i^\dag \quad (\text{where} \sum_i p_i =1 ).
\end{align*}

\subsection{Nonlocal correlations with restricted operations}\label{sec:Nonlocal correlations with restricted operations}
Here we briefly motivate why a non-stabilizer operation is necessary to exhibit non-locality. Consider the maximally entangled state $\ket{\Phi}=\tfrac{\ket{00}+\ket{11}}{\sqrt{2}}$, then its Pauli expectation values, defined as
\begin{align*}
\text{JK}=\Tr\left(\ketbra{\Phi}{\Phi} \sigma_j \otimes \sigma_k\right)  \quad j,k \in \{x,y,z\}
\end{align*}
are all zero except for
\begin{align*}
\text{XX}=-\text{YY}=\text{ZZ}=1
\end{align*}
The Pauli expectation values for $\ket{\Phi}$ can be recreated exactly by two spatially separated parties \textbf{A} and \textbf{B}, given that they share $3$ random (unbiased) bits $\{r_1,r_2,r_3\}\in\{0,1\}$, and they both obey the following set of rules \cite{dabacon}
\begin{enumerate}
\item Measurement in the $X$ direction $\leftrightarrow$ \textbf{A} and \textbf{B} both output $-1^{r_1}$
\item Measurement in the $Y$ direction $\leftrightarrow$ \textbf{A} outputs $-1^{r_2}$, \textbf{B} outputs $-1^{r_2+1}$
\item Measurement in the $Z$ direction $\leftrightarrow$ \textbf{A} and \textbf{B} both output $-1^{r_3}$
\end{enumerate}
A little thought shows that a similar scheme would work for any bipartite entangled state created by stabilizer operations. That such a scheme suffices, using only shared randomness, indicates that no purely quantum mechanical effects are needed to describe such an experiment.

\section{Bell Inequalities}\label{sec:Bell Inequalities}

\begin{figure}[ht]
\begin{center}
\includegraphics[scale=1.1,bb=0 0 200 110]{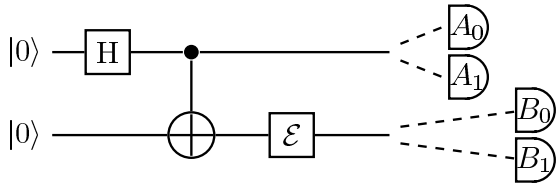}
\end{center}
\caption{\label{fig:tikz_circuit_ising_anyons} A simple set-up to detect nonlocality: Here, two possible measurement settings for the first (second) qubit are denoted $A_i$ ($B_j$). When $A_i$ and $B_j$ are constrained to be Pauli operators, then $\mathcal{E}$ must be a non-stabilizer operation if any Bell inequality (e.g., that given in \eqref{eqn:CHSH}) is to be violated.}
\end{figure}

We will attempt to perform an experiment that, unlike the example of Section \ref{sec:Nonlocal correlations with restricted operations}, is not describable by local hidden variables. The basic setup is depicted in Figure \ref{fig:tikz_circuit_ising_anyons}, where $\mathcal{E}$ is applied to one half of a maximally entangled state $\protect{(\ket{00}+\ket{11})/\sqrt{2}}$ and then Pauli measurements are performed. We can allow for either two or three distinct Pauli measurements to be performed on each qubit (Figure \ref{fig:tikz_circuit_ising_anyons} depicts the scenario in which only two measurements are performed on each side), and these lead to different types of Bell inequality.

For a given bipartite state $\varrho$, and using a notation similar to that of Collins and Gisin \cite{Collins2004}, we can arrange the expectation values for a given nonlocality experiment with Pauli measurements in a table i.e.,
\begin{align}
\left(
\begin{array}{cccc}
 \text{II} & \text{XI} & \text{YI} & \text{ZI} \\
 \text{IX} & \text{XX} & \text{YX} & \text{ZX} \\
 \text{IY} & \text{XY} & \text{YY} & \text{ZY} \\
 \text{IZ} & \text{XZ} & \text{YZ} & \text{ZZ}
\end{array}
\right) \label{eqn:CGtable}
\end{align}
where XY is the expectation value $\Tr(\varrho \sigma_x \otimes \sigma_y)$ and so on. Since each Pauli operator has eigenvalues $\pm 1$, we can list all possible matrices of the form \eqref{eqn:CGtable} that correspond to deterministic local configurations. A deterministic local configuration is one in which the local Pauli expectation values are extremal and, when multiplied, provide the nonlocal Pauli expectation value (e.g., where identities like (XI)(IY)=XY hold, with XI,IY,XY all either $\pm 1$). By letting $a,b,\ldots,f \in \{0,1\}$ take on all $2^6$ assignments, then the following matrix fully describes all possible local configurations,
\begin{align}
\left(
\begin{array}{cccc}
 1 & (-1)^a & (-1)^b & (-1)^c \\
 (-1)^d & (-1)^{a+d} & (-1)^{b+d} & (-1)^{c+d} \\
 (-1)^e & (-1)^{a+e} & (-1)^{b+e} & (-1)^{c+e} \\
 (-1)^f & (-1)^{a+f} & (-1)^{b+f} & (-1)^{c+f}
\end{array}
\right). \label{eqn:LocalConfigs}
\end{align}
If an experimentalist's measured expectation values correspond \emph{exactly} to any of the above local configurations, then these expectation values do not exhibit any nonlocality. Clearly, any realistic experiment will not obey such a strict condition and the question of exhibiting nonlocality becomes richer. The experimentalist must now check that the measured expectation values cannot be expressed as a probabilistic combination of local configurations. This naturally leads to the notion of convex geometry and bounded polyhedra in higher dimensions (these are usually called polytopes).

Ignoring the constant term in the top-left, the $2^6$ matrices in \eqref{eqn:LocalConfigs} can each be identified with a vector in $\mathbb{R}^{15}$. These local configuration vectors are called vertices and the convex hull of these vertices (i.e. the set of vectors that is expressible as a probabilistic combination of the vertices) describes a polytope in $\mathbb{R}^{15}$ (this is analagous to how the 8 vertices $(\pm1,\pm1,\pm1)$ describe a solid cube in $\mathbb{R}^3$). The interior of this polytope describes all possible LHV models. Conversely, any table of expectation values \eqref{eqn:CGtable} that does not lie inside this polytope exhibits genuine quantum nonlocality. The usual, textbook, way of identifying nonlocality is via Bell inequalities like
\begin{align}
\text{XX}+\text{XY}+\text{YX}-\text{YY} \leq 2. \label{eqn:CHSH}
\end{align}
In fact, (tight) Bell inequalities such as these are actually the defining inequalities for the bounding faces (facets) of the LHV polytope that we have just described. Software such as Avis' lrs \cite{Avis2000Revised} can be used to derive all the facets of LHV polytope, given the $2^6$ local configuration vectors in \eqref{eqn:LocalConfigs} as input (and vice versa). In order to unify our notation, we rearrange the inequality \eqref{eqn:CHSH} and  then rewrite in table form
\begin{align}
&2-\text{XX}-\text{XY}-\text{YX}+\text{YY}\geq 0 \\
\Rightarrow  &\left(
\begin{array}{cccc}
 2 & 0 & 0 & 0 \\
 0 & -1 & -1 & 0 \\
 0 & -1 & 1 & 0 \\
 0 & 0 & 0 & 0
\end{array}
\right) \cdot  \left(
\begin{array}{cccc}
 \text{II} & \text{XI} & \text{YI} & \text{ZI} \\
 \text{IX} & \text{XX} & \text{YX} & \text{ZX} \\
 \text{IY} & \text{XY} & \text{YY} & \text{ZY} \\
 \text{IZ} & \text{XZ} & \text{YZ} & \text{ZZ}
\end{array}
\right) \geq 0 \label{eqn:BIasFacet}
\end{align}
where the dot product between two matrices $M,N$ behaves like the familiar vector dot product i.e., $M \cdot N = \Tr(M^T N)=\sum_{i,j} M_{i,j}N_{i,j}$. The full list of bounding inequalities for the LHV polytope is described in a concise form in Section \ref{sec:Results}.

\section{Non-stabilizer operations}
In the previous section, ascertaining whether measurement results had an LHV model amounted to the question of whether the results could be expressed as a probabilistic combination of local configurations. In the context of UQC, the relevant question is now whether $\mathcal{E}$ is expressible as a probabilistic combination of Clifford gates; for example, an operation $\mathcal{E}$ given by
\begin{align*}
\mathcal{E}(\rho)=\frac{1}{3}H\rho H^\dag +\frac{1}{4}S\rho S^\dag +\frac{5}{12}\rho
\end{align*}
is clearly of no use in the quest to achieve UQC. The general prerequisite for $\mathcal{E}$ to be useful is
\begin{align}
\mathcal{E}(\rho) \neq \sum_{i=1}^{24} p_iC_i \rho C_i^\dag \quad \left(\text{where }\sum_{i=1}^{24} p_i =1\right)  \label{eqn:NonCliffOps}
\end{align}
If we use the \Jam isomorphism (as defined in \eqref{eqn:JamIso}) to represent operations, then a completely equivalent condition to \eqref{eqn:NonCliffOps} is given by
\begin{align}
\vre \neq \sum_{i=1}^{24} p_i \ket{J_{C_i}}\bra{J_{C_i}} \quad \left(\text{where }\sum_{i=1}^{24} p_i =1 \text{ and } \ket{J_{C_i}}=\left(\mathbb{I}\otimes C_i\right) \ket{\Phi} \right)
\end{align}
To test whether an unknown operation satisfies the above requirement, one is naturally lead to the concept of convex polytopes once more. The $24$ Clifford gates, when represented as $\ket{J_{C_i}}\bra{J_{C_i}}$ and decomposed in the Pauli basis, form the $24$ vertices of what is called the Clifford polytope. In fact, the Clifford operations only span a 9-dimensional subspace of $\mathbb{R}^{15}$ because the six coefficients IX,IY,IZ,XI,YI and ZI are all identically zero when $\mathcal{E}(\mathbb{I})=\mathbb{I}$. The Clifford gates given in \eqref{eqn:HandS}, for example, correspond to the following two vertices
\begin{align*}
v_{H}=\left(
\begin{array}{cccc}
 1 & 0 & 0 & 0 \\
 0 & 0 & 0 & 1 \\
 0 & 0 & 1 & 0 \\
 0 & 1 & 0 & 0
\end{array}
\right) \quad v_{S}=\left(
\begin{array}{cccc}
 1 & 0 & 0 & 0 \\
 0 & 0 & 1 & 0 \\
 0 & 1 & 0 & 0 \\
 0 & 0 & 0 & 1
\end{array}
\right)
\end{align*}

\begin{figure}[ht]
\begin{center}
\includegraphics[scale=1.1,bb=0 0 200 110]{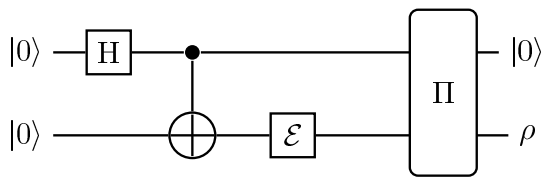}
\caption{\label{fig:rho_from_op} A circuit to help achieve universal quantum computation via magic state distillation:  Every element of this circuit except $\mathcal{E}$ is implementable using stabilizer operations. When $\mathcal{E}$ exhibits nonlocality in the setup of Figure \ref{fig:tikz_circuit_ising_anyons}, then $\mathcal{E}$ used in the above circuit produces ancillas $\rho$ that are useful for magic state distillation subroutine (MSD circuit not depicted). The block containing $\Pi$ stands for a two-qubit Pauli measurement (e.g.~parity measurement) wherein we postselect on the desired outcome. }
\end{center}
\end{figure}

The facets of the Clifford polytope \cite{Buhrmanetal:2006,WvDMH:2009,WvDMH:2010} are bounding inequalities that partition the set of all valid \vre into those that represent $\mathcal{E}$ satisfying \eqref{eqn:NonCliffOps} and those that do not. In other words, any operation $\mathcal{E}$ that satisfies \eqref{eqn:NonCliffOps} violates (at least) one of these facet inequalities. These inequalities, which we label $I$, can be cast in matrix form, in a completely equivalent way to what was done in \eqref{eqn:BIasFacet}, and it turns out there are two distinct classes of facet \cite{Buhrmanetal:2006} -- a representative example of each of which we provide below.
\begin{align}
&\mathcal{I}_{\alpha}=\{\text{all facets of type } I_{\alpha} \text{ or } I_{\alpha}^T\},\quad &I_{\alpha}=\left(
\begin{array}{cccc}
 1 & 0 & 0 & 0 \\
 0 & 1 & 0 & 0 \\
 0 & 1 & 0 & 0 \\
 0 & 1 & 0 & 0
\end{array}
\right)  \\
&\mathcal{I}_{\beta}=\{\text{all facets of type } I_{\beta} \},\quad &I_{\beta}=\left(
\begin{array}{cccc}
 1 & 0 & 0 & 0 \\
 0 & -1 & -1 & 0 \\
 0 & -1 & 1 & 0 \\
 0 & 0 & 0 & 1
\end{array}
\right) \label{eqn:Bdefn}
\end{align}
The total number of facets, required to completely characterize the boundary of the Clifford polytope, is $120$, which can be broken down into $|\mathcal{I}_\alpha|+|\mathcal{I}_\beta|=(24+24)+72$. We are primarily interested in facets from $\mathcal{I}_\beta$ because it was shown in \cite{WvDMH:2009} that an operation violating such a facet can always be used to create an ancilla $\rho$ that is useful for magic state distillation. The circuit to create $\rho$ is given in Figure \ref{fig:rho_from_op}, where the choice of which two-qubit Pauli measurement $\Pi$ to perform is determined by the particular facet that is violated by $\mathcal{E}$.

\section{Results}\label{sec:Results}

A full facet description of the LHV polytope as described in Section \ref{sec:Bell Inequalities} comprises 684 distinct facets, and this facet description can partitioned into 3 different classes i.e,
\begin{align}
\mathcal{I}_{triv}=\{\text{all facets of type } I_{triv}\},\quad I_{triv}=\left(
  \begin{array}{cccc}
    1 & -1 & 0 & 0 \\
    -1 & 1 & 0 & 0 \\
    0 & 0 & 0 & 0 \\
    0 & 0 & 0 & 0 \\
  \end{array}
\right) \label{eqn:triv} \\ \mathcal{I}_{2222}=\{\text{all facets of type } I_{2222}\},\quad  I_{2222}=\left(
\begin{array}{cccc}
 2 & 0 & 0 & 0 \\
 0 & -1 & -1 & 0 \\
 0 & -1 & 1 & 0 \\
 0 & 0 & 0 & 0
\end{array}
\right) \label{eqn:I222} \\ \mathcal{I}_{3322}=\{\text{all facets of type } I_{3322}\},\quad  I_{3322}=\left(
  \begin{array}{cccc}
    4 & -1 & -1 & 0 \\
    -1 & 1 & 1 & -1 \\
    -1 & 1 & 1 & 1 \\
    0 & -1 & 1 & 0 \\
  \end{array}
\right)
\end{align}
The number of distinct facets in each class is $|\mathcal{I}_{triv}|=36$, $|\mathcal{I}_{2222}|=72$ and $\protect{|\mathcal{I}_{3322}|=576}$. Inequalities of the type $I_{triv}$ are considered trivial, since the restrictions they impose are satisfied by any bipartite quantum state $\varrho$. The representative inequality $I_{triv}$ given in \eqref{eqn:triv} amounts to $\protect{\Tr\left[(\mathbb{I}-\sigma_x)\otimes (\mathbb{I}-\sigma_x) \varrho\right]\geq 0}$, for example. Inequalities of the type $I_{2222}$ are usually known as Clauser-Horne-Shimony-Holt (CHSH) inequalities \cite{Clauser1969} and an example of one such inequality was already discussed in the context of Equation \eqref{eqn:BIasFacet}. These $I_{2222}$ inequalities appear, in a different form closer to that of \eqref{eqn:CHSH}, in most textbook accounts of nonlocality (see e.g. \cite{NielsenChuang:2000}). Bell inequalities of the type $I_{3322}$ are less well known but have a long history, appearing as early as 1981 \cite{Froissart} and rediscovered more recently in \cite{Pitowsky2001,Sliwa2003,Collins2004}. Finally we note the reasoning behind the notation $I_{ccdd}$; the numbers $ccdd$ in the subscript refer to the number of measurement settings, $c$, available to each party, wherein each measurement has a number, $d$, of possible outcomes (e.g., inequalities of the type $I_{3322}$ can only arise when there are $3$ possible measurement settings on each side of the bipartite state).

The $I_{2222}$ inequalities require only two Pauli measurements on each side whereas $I_{3322}$ inequalities require three on each side. It turns out that the $I_{2222}$ inequalities are sufficient for our purposes. In fact, $I_{3322}$ inequalities are irrelevant (redundant) for a very large subset of all possible quantum operations $\mathcal{E}$. If one makes the assumption that $\mathcal{E}(\mathbb{I})=\mathbb{I}$, which is the case for many of the most important noise models, then local expectation values for $\varrho$ are identically zero, and coefficients of IX,IY,IZ,XI,YI and ZI in $I_{3322}$ facets can be ignored. With this modification, the $I_{3322}$ inequalities are expressible as linear combinations of $I_{2222}$ (CHSH) inequalities, and hence the $I_{3322}$ inequalities are irrelevant. One can check that all (purely nonlocal) $I_{3322}$ inequalities decompose into four $I_{2222}$ inequalities, as in the following example
\begin{align}
\left(
  \begin{array}{cccc}
    4 & 0 & 0 & 0 \\
    0 & 1 & 1 & -1 \\
    0 & 1 & 1 & 1 \\
    0 & -1 & 1 & 0 \\
  \end{array}
\right)=
&\frac{1}{2}\left[\left(
  \begin{array}{cccc}
    2 & 0 & 0 & 0 \\
    0 & 0 & 0 & 0 \\
    0 & 1 & 1 & 0 \\
    0 & -1 & 1 & 0 \\
  \end{array}
\right)+\left(
  \begin{array}{cccc}
    2 & 0 & 0 & 0 \\
    0 & 0 & 1 & -1 \\
    0 & 0 & 1 & 1 \\
    0 & 0 & 0 & 0 \\
  \end{array}
\right)\right.\\&+\left.\left(
  \begin{array}{cccc}
    2 & 0 & 0 & 0 \\
    0 & 1 & 0 & -1 \\
    0 & 1 & 0 & 1 \\
    0 & 0 & 0 & 0 \\
  \end{array}
\right)+\left(
  \begin{array}{cccc}
    2 & 0 & 0 & 0 \\
    0 & 1 & 1 & 0 \\
    0 & 0 & 0 & 0 \\
    0 & -1 & 1 & 0 \\
  \end{array}
\right)\right] \nonumber
\end{align}

%\begin{align*}
% \left(
%\begin{array}{cccc}
% 1 & 0 & 0 & 0 \\
% 0 & -1 & -1 & 0 \\
% 0 & -1 & 1 & 0 \\
% 0 & 0 & 0 & 1
%\end{array}
%\right),  \left(
%\begin{array}{cccc}
% 2 & 0 & 0 & 0 \\
% 0 & -1 & -1 & 0 \\
% 0 & -1 & 1 & 0 \\
% 0 & 0 & 0 & 0
%\end{array}
%\right)
%\end{align*}

We are now ready to state the main result:
\begin{theorem}\label{thm:theorem1}
If there is an operation $\mathcal{E}$ violating a facet from $\mathcal{I}_{2222}$ \eqref{eqn:I222} (i.e., violating a CHSH inequality), then $\mathcal{E}$ also violates a facet (of the Clifford polytope) from $\mathcal{I}_{\beta}$ \eqref{eqn:Bdefn}. Using results from \cite{WvDMH:2009}, such $\mathcal{E}$ enable UQC (via magic state distillation) when supplemented with topologically protected operations.
\end{theorem}

Mathematically, we have the following statement
\begin{align}\label{eqn:theoreminmath}
&\forall \text{ physically valid }\mathcal{E},\quad \forall\ I_{2222}\in \mathcal{I}_{2222},\quad  \exists\ I_{\beta} \in \mathcal{I}_{\beta} \text{ such that }\\
&\vre \text{ violates } I_{\beta} \text{ at least as much as it violates } I_{2222} \nonumber
\end{align}
Without loss of generality, we assume that the canonical representative, $I_{2222}$ \eqref{eqn:I222}, of $\mathcal{I}_{2222}$ is violated, and show that $I_{\beta}$ \eqref{eqn:Bdefn} from $\mathcal{I}_{\beta}$ is necessarily also violated. This is a consequence of the following proposition, which is easily proved
%\begin{widetext}
\begin{align}\label{eqn:keyinequality}
 \left(
\begin{smallmatrix}
 1 & 0 & 0 & 0 \\
 0 & -1 & -1 & 0 \\
 0 & -1 & 1 & 0 \\
 0 & 0 & 0 & 1
\end{smallmatrix}
\right)\cdot \left(
\begin{smallmatrix}
 \text{II} & \text{XI} & \text{YI} & \text{ZI} \\
 \text{IX} & \text{XX} & \text{YX} & \text{ZX} \\
 \text{IY} & \text{XY} & \text{YY} & \text{ZY} \\
 \text{IZ} & \text{XZ} & \text{YZ} & \text{ZZ}
\end{smallmatrix}
\right) \leq \left(
\begin{smallmatrix}
 2 & 0 & 0 & 0 \\
 0 & -1 & -1 & 0 \\
 0 & -1 & 1 & 0 \\
 0 & 0 & 0 & 0
\end{smallmatrix}
\right) \cdot \left(
\begin{smallmatrix}
 \text{II} & \text{XI} & \text{YI} & \text{ZI} \\
 \text{IX} & \text{XX} & \text{YX} & \text{ZX} \\
 \text{IY} & \text{XY} & \text{YY} & \text{ZY} \\
 \text{IZ} & \text{XZ} & \text{YZ} & \text{ZZ}
\end{smallmatrix}
\right).
\end{align}
%\end{widetext}
Rearranging the above inequality gives
\begin{align}
\left(
\begin{array}{cccc}
 -1 & 0 & 0 & 0 \\
 0 & 0 & 0 & 0 \\
 0 & 0 & 0 & 0 \\
 0 & 0 & 0 & 1
\end{array}
\right)\cdot \left(
\begin{array}{cccc}
 \text{II} & \text{XI} & \text{YI} & \text{ZI} \\
 \text{IX} & \text{XX} & \text{YX} & \text{ZX} \\
 \text{IY} & \text{XY} & \text{YY} & \text{ZY} \\
 \text{IZ} & \text{XZ} & \text{YZ} & \text{ZZ}
\end{array}
\right) \leq 0
\end{align}
which simply stands for
\begin{align}\label{eqn:tightcondition}
-1+\text{ZZ}\leq0.
\end{align}
This is always satisfied since Pauli expectation values are in the range $[-1,1]$.
To create an ancilla for magic state distillation (and consequently enable UQC), the best measurement $\Pi$ to use in the circuit of Figure \ref{fig:rho_from_op} is the parity measurement postselcted on even parity i.e., $\Pi=\frac{1}{2}\left(\mathbb{I}+\sigma_z\sigma_z\right)$.

Note that $|\mathcal{I}_\beta|=|\mathcal{I}_{2222}|=72$. Examining the matrix representation of these sets of facets, one sees that each $I_\beta\in \mathcal{I}_\beta$ can be paired one-to-one with each $I_{2222} \in \mathcal{I}_{2222}$, by matching facets whose non-identity coefficients differ in only one position (e.g., the coefficient of ZZ in our canonical $I_{2222}$ \eqref{eqn:I222} is $0$, whereas for our canonical $I_\beta$ \eqref{eqn:Bdefn} the coefficient is $+1$). This provides a recipe for identifying the relevant $I_\beta$ such that \eqref{eqn:theoreminmath} holds, for a given $I_{2222}$ (the CHSH inequality that we presume has been violated). Any such pair $\{I_{2222},I_\beta\}$ will obey an inequality of the form \eqref{eqn:keyinequality}. The general expression, analagous to \eqref{eqn:tightcondition}, for different facets from $\mathcal{I}_{2222}$ and suitably chosen facets from $\mathcal{I}_\beta$ is
\begin{align}
-1\pm \Tr(\sigma_j\otimes \sigma_k \vre) \leq 0 \quad j,k \in \{x,y,z\}
\end{align}
and this clearly always holds. In order to use $\mathcal{E}$ to obtain ancillas useful for magic state distillation, a suitable measurement $\Pi$ in Figure \ref{fig:rho_from_op} includes one of the form $\Pi=\frac{1}{2}\left(\mathbb{I}\pm\sigma_j\sigma_k\right)$ \cite{WvDMH:2009}.

\subsection{Example: Phase Gate subject to Dephasing}\label{sec:Example: Phase Gate subject to Dephasing}

The phase gate, $U_z(\theta)$, defined as
\begin{align*}
U_z(\theta)=\left(
                                                   \begin{array}{cc}
                                                     1 & 0 \\
                                                     0 & e^{i \theta} \\
                                                   \end{array}
                                                 \right)=\left(
                                                   \begin{array}{cc}
                                                     e^{-i \frac{\theta}{2}} & 0 \\
                                                     0 & e^{i \frac{\theta}{2}} \\
                                                   \end{array}
                                                 \right)
\end{align*}
is an element of the Clifford group when $\theta$ is a multiple of $\pi/2$. Whilst technically any non-Clifford angle $\theta$ would suffice to enable UQC, one typically seeks to implement the so called ``pi-over-eight'' gate $U_z(\tfrac{\pi}{4})$ because it possesses additional desirable properties \cite{Boykin2000}. For the case of Ising anyons, it has been proposed \cite{Bonderson2010a,Clarke2010} to perform a phase gate using a sack-like geometry, wherein the sack contains a single qubit, and an anyonic edge current can either follow the exterior boundary of the sack or tunnel across the constriction of the sack. In \cite{Bonderson2010a} it was shown that, subject to certain assumptions, the overall evolution of the qubit state can be described by a superoperator $\mathcal{E}(\rho(t_0))=\rho(t)$ where the matrix elements of the density operator change like
\begin{align*}
\rho(t)=\left(
          \begin{array}{cc}
            \rho_{00}(t_0) & e^{-\frac{s^2}{2}}e^{-i \theta}\rho_{01}(t_0) \\
            e^{-\frac{s^2}{2}}e^{i \theta}\rho_{10}(t_0) & \rho_{11}(t_0) \\
          \end{array}
        \right)
\end{align*}
or, equivalently in the Kraus operator form,
\begin{align}
       &\rho(t)=E_0 \rho(t_0) E_0^\dag+E_1 \rho(t_0) E_1^\dag \quad \text{where}\label{eqn:PHYSkrauss}\\
        &E_0=\sqrt{\frac{1+e^{-\frac{s^2}{2}}}{2}}\left(
                                                   \begin{array}{cc}
                                                     1 & 0 \\
                                                     0 & e^{i \theta} \\
                                                   \end{array}
                                                 \right) \qquad         E_1=\sqrt{\frac{1-e^{-\frac{s^2}{2}}}{2}}\left(
                                                   \begin{array}{cc}
                                                     1 & 0 \\
                                                     0 & -e^{i \theta} \\
                                                   \end{array}
                                                 \right)\nonumber
\end{align}
As a side note; in quantum information theory, the description of this process in terms of Kraus operators would probably be rewritten
\begin{align}
&\mathcal{E}(\rho)=E_0 \rho E_0^\dag+E_1 \rho E_1^\dag \text{ with } \label{eqn:QIkrauss}\\
&E_0=\sqrt{1-p}\ U_z(\theta), \quad E_1=\sqrt{p}\ \sigma_z U_z(\theta),\quad \text{ where } p=\frac{1-e^{-\frac{s^2}{2}}}{2} \nonumber
\end{align}
which is completely equivalent. The natural interpretation is that with probability $1-p$ the desired $U_z(\theta)$ gate is performed, and with probability $p$ an undesired rotation $\sigma_z U_z(\theta)$ is performed. This undesired rotation is actually the worst possible noise that could be inflicted \cite{VirmaniHuelgaPlenio:2005} (i.e., it requires the smallest noise rate, $p$, to make the overall $\mathcal{E}$ expressible as a probabilistic combination of Clifford gates).

The operation, $\mathcal{E}$, of \eqref{eqn:PHYSkrauss}, when expressed as a \Jam state, \vre, and decomposed in the Pauli basis, takes the form
\begin{align}
 \left(
\begin{array}{cccc}
 \text{II} & \text{XI} & \text{YI} & \text{ZI} \\
 \text{IX} & \text{XX} & \text{YX} & \text{ZX} \\
 \text{IY} & \text{XY} & \text{YY} & \text{ZY} \\
 \text{IZ} & \text{XZ} & \text{YZ} & \text{ZZ}
\end{array}
\right)=\left(
\begin{array}{cccc}
 1 & 0 & 0 & 0 \\
 0 & e^{-\frac{s^2}{2}} \text{Cos}[\theta ] & e^{-\frac{s^2}{2}} \text{Sin}[\theta ] & 0 \\
 0 & e^{-\frac{s^2}{2}} \text{Sin}[\theta ] & -e^{-\frac{s^2}{2}} \text{Cos}[\theta ] & 0 \\
 0 & 0 & 0 & 1
\end{array}
\right) \label{eqn:DephasinginPB}
\end{align}
and so the Bell inequality given in \eqref{eqn:BIasFacet} reads
\begin{align*}
&\text{XX}+\text{XY}+\text{YX}-\text{YY} \leq 2
\Rightarrow &e^{-\frac{s^2}{2}}\left([\cos \theta] +[\sin \theta] +[\sin \theta] - [-\cos \theta] \right)\leq 2
\end{align*}
For the optimal choice of angle, $\theta=\tfrac{\pi}{4}$, this inequality is violated for the parameter range  $0\leq s < \sqrt{\ln 2}$. In terms of the simplified Kraus operators given in \eqref{eqn:QIkrauss} the range of allowable noise rates $p$, while still giving CHSH violation is given by $0\leq p \lesssim 14\% $.

Another simple calculation shows that the facet $I_\beta$ is also violated (hence UQC is possible) for the same noise rates $0\leq p \lesssim 14\% $. One can actually see this straight away by noting that \eqref{eqn:DephasinginPB} implies that $-1+$ZZ$=0$, which means (with reference to \eqref{eqn:tightcondition}) that the threshold noise rate is the same for violation of $I_\beta$ and $I_{2222}$. This will also be true for any noise model that is expressible as a probabilistic combination of rotations about the $z$-axis (and similarly for operations comprised of rotations about $x$ and $y$ axes too).

The noise rates for which Bell inequality violation is possible, and the noise rates for which UQC is possible, generally do not coincide. For example, a  $U_z(\tfrac{\pi}{4})$ gate under depolarizing noise, modeled as
\begin{align*}
\mathcal{E}(\rho)=(1-p) U_z(\tfrac{\pi}{4}) \rho U_z(\tfrac{\pi}{4})^\dag+p\frac{\mathbb{I}}{2}
\end{align*}
enables UQC, but does not violate any Bell inequality, for the parameter range $0.29\lesssim p \lesssim 0.45$. In the current context, a non-Clifford operation $\mathcal{E}$ is necessary if one hopes to observe nonlocality (as shown in Section \ref{sec:Nonlocal correlations with restricted operations}), but not always sufficient. However, if we expand the problem to allow multiple (possibly simultaneous) uses of $\mathcal{E}$, in a more complex circuit than that of Figure \ref{fig:tikz_circuit_ising_anyons}, then it is an interesting question whether $\mathcal{E}$ being non-Clifford could be both a necessary and sufficient condition for the violation of a bipartite Bell inequality.

\section{Conclusion}
By using techniques from convex geometry, we can show that detection of nonlocality in the encoded qubits of an Ising anyon TQC serves as a benchmark for universal quantum computation. The strength of this approach is that it does not consider specific noise models, but rather considers general quantum operations. Given an unknown operation that has been observed to enable violation of a CHSH inequality (using two distinct Pauli measurements on each half of an entangled state), we can prescribe a way of using this operation to manufacture ancillas that are suitable for magic state distillation. When used in conjunction with topologically protected stabilizer operations, ancillas of this kind allow for universal quantum computation.

\section{Acknowledgements}
We thank Earl Campbell for helpful comments on a previous version of this manuscript. MH was supported by an Empower Fellowship from the Irish Research Council for Science, Engineering and Technology. JV acknowledges funding from Science Foundation Ireland under grant SFI 10/IN.1/I3013.


\begin{thebibliography}{99}


%1
\bibitem{Moore1991}
Moore, G and Read, N,
\newblock ``Nonabelions in the fractional quantum Hall effect'',
\newblock Nucl.~Phys.~B \textbf{360}, 362--396, (1991).
\href{http://dx.doi.org/10.1016/0550-3213(91)90407-O}{http://dx.doi.org/10.1016/0550-3213(91)90407-O}

%2
\bibitem{Bonderson2006}
Bonderson, P and Kitaev, A
\newblock ``Detecting non-Abelian statistics in the $\nu= 5/2$ fractional quantum Hall state'',
\newblock Phys.~Rev.~Lett.~ \textbf{96}, 016803, (2006).
\href{http://link.aps.org/doi/10.1103/PhysRevLett.96.016803}{http://link.aps.org/doi/10.1103/PhysRevLett.96.016803}

%3
\bibitem{Alicea2011}
Alicea, J and Oreg, Y and Refael, G and von Oppen, F
\newblock ``Non-Abelian statistics and topological quantum information processing in 1D wire networks'',
\newblock Nature Physics \textbf{7}, 412--417 (2011).
\href{http://dx.doi.org/10.1038/nphys1915}{http://dx.doi.org/10.1038/nphys1915}

%4
\bibitem{Ivanov2001}
Ivanov, D. a.
\newblock ``Non-Abelian Statistics of Half-Quantum Vortices in p-Wave Superconductors'',
\newblock Phys.~Rev.~Lett.~ \textbf{86}, 268, (2001).
\href{http://link.aps.org/doi/10.1103/PhysRevLett.86.268}{http://link.aps.org/doi/10.1103/PhysRevLett.86.268}


%5
\bibitem{Kitaev2006a}
Kitaev, A.
\newblock ``Anyons in an exactly solved model and beyond'',
\newblock Annals of Physics \textbf{321}, 1, 2--111, (2006).
\href{http://dx.doi.org/10.1016/j.aop.2005.10.005}{http://dx.doi.org/10.1016/j.aop.2005.10.005}

%6
 \bibitem{Plenio2010}
M.~B.~Plenio and S.~Virmani,
``Upper bounds on fault tolerance thresholds of noisy Clifford-based quantum computers''
\newblock New Journal of Physics \textbf{12}, number 3, 033012 , (2010).
\href{http://dx.doi.org/10.1088/1367-2630/12/3/033012}{http://dx.doi.org/10.1088/1367-2630/12/3/033012}


%7
\bibitem{Bravyi2006}
Bravyi, Sergey,
\newblock ``Universal quantum computation with the $\nu=\tfrac{5}{2}$ fractional quantum Hall state'',
\newblock Phys.~Rev.~A \textbf{73}, 042313, (2006).
\href{http://link.aps.org/doi/10.1103/PhysRevA.73.042313}{http://link.aps.org/doi/10.1103/PhysRevA.73.042313}

%8
\bibitem{Georgiev:2006}
L.~S.~Georgiev,
``Topologically protected gates for quantum computation with
  non-abelian anyons in the Pfaffian quantum Hall state'',
\newblock Phys.~Rev.~B \textbf{74}, 235112 (2006)''

%9
\bibitem{Bonderson201x}
P. Bonderson \emph{et al.}, In preparation.


%10
\bibitem{NielsenChuang:2000}
M.~A.~Nielsen and I.~L.~Chuang,
\newblock {\em Quantum Computation and Quantum Information}.
\newblock Cambridge University Press, Cambridge, (2000).

%11
\bibitem{Buhrmanetal:2006}
 H.~Buhrman, R.~Cleve, M.~Laurent, N.~Linden, A.~Schrijver and F.~Unger,
``New Limits on Fault-Tolerant Quantum Computation'',
 \emph{Annual IEEE Symposium on Foundations of Computer Science,}
 pp.~411--419, (2006).
 \href{http://dx.doi.org/10.1109/FOCS.2006.50}{http://dx.doi.org/10.1109/FOCS.2006.50}

%12
\bibitem{VirmaniHuelgaPlenio:2005}
S.~Virmani and S~F.~Huelga and M~B.~Plenio,
\newblock Classical simulability, entanglement breaking, and quantum computation thresholds,
Phys.~Rev.~A \textbf{71}, 042328 (2005).
\href{http://link.aps.org/doi/10.1103/PhysRevA.71.042328}{http://link.aps.org/doi/10.1103/PhysRevA.71.042328}

%13
 \bibitem{WvDMH:2009}
W.~van~Dam and M.~Howard,
``Tight Noise Thresholds for Quantum Computation with Perfect Stabilizer Operations''
\newblock Phys.~Rev.~Lett. \textbf{103}, 170504, (2009).
\href{http://dx.doi.org/10.1103/PhysRevLett.103.170504}{http://dx.doi.org/10.1103/PhysRevLett.103.170504}


%13.5
\bibitem{Knill05}
E.~Knill ,
``Quantum Computing with Realistically Noisy Devices"
\newblock Nature \textbf{434}, pp.~39--44, (2005).
\href{http://dx.doi.org/10.1038/nature03350}{http://dx.doi.org/10.1038/nature03350}

%13.6
\bibitem{BravyiKitaev:2005}
S.~Bravyi and A.~Kitaev,
`` Universal quantum computation with ideal Clifford gates and noisy ancillas'',
\newblock Phys.~Rev.~A \textbf{71}, 022316 (2005).
\href{http://dx.doi.org/10.1103/PhysRevA.71.022316}{http://dx.doi.org/10.1103/PhysRevA.71.022316}

%13.7
\bibitem{ReichardtMagic09}
B.~Reichardt,
``Quantum universality by state distillation "
\newblock Quantum Inf. Comput. \textbf{9},  pp.~1030--1052 (2009).
\href{http://dl.acm.org/citation.cfm?id=2012105}{http://dl.acm.org/citation.cfm?id=2012105}

%13.8
\bibitem{CampbellBrowne:2009}
E.~T.~Campbell and D.~E.~Browne,
``On the Structure of Protocols for Magic State Distillation'',
in \emph{Theory of Quantum Computation, Communication, and Cryptography}, (editor: Childs, Andrew and Mosca, Michele),
\emph{Lecture Notes in Computer Science,} Volume 5906, (Springer Berlin / Heidelberg, 2009), pp.~20--32.
\href{http://dx.doi.org/10.1007/978-3-642-10698-9_3}{http://dx.doi.org/10.1007/978-3-642-10698-9\_3}

%13.9
\bibitem{CampbellBrowne:2010}
E.~T.~Campbell and D.~E.~Browne,
``Bound States for Magic State Distillation in Fault-Tolerant Quantum Computation"
\newblock Phys.~Rev.~Lett. \textbf{104}, 030503, (2010).
\href{}{}


%14
 \bibitem{Brennen2009}
Brennen, G K and Iblisdir, S and Pachos, J K and Slingerland, J K,
``Non-locality of non-Abelian anyons''
\newblock New Journal of Physics \textbf{10}, 103023, (2009).
\href{http://dx.doi.org/10.1088/1367-2630/11/10/103023}{http://dx.doi.org/10.1088/1367-2630/11/10/103023}

%15
 \bibitem{Deng2010}
Deng, Dong-Ling and Wu, Chunfeng and Chen, Jing-Ling and Oh, C.,
``Fault-Tolerant Greenberger-Horne-Zeilinger Paradox Based on Non-Abelian Anyons''
\newblock Phys.~Rev.~Lett. \textbf{105}, 060402, (2010).
\href{http://dx.doi.org/10.1103/PhysRevLett.105.060402}{http://dx.doi.org/10.1103/PhysRevLett.105.060402}


%16
\bibitem{Ratanje2011}
Ratanje, N. and Virmani, S.l,
\newblock ``Generalized state spaces and nonlocality in fault-tolerant quantum-computing schemes'',
\newblock Phys.~Rev.~A \textbf{83}, 032309, (2011).
\href{http://link.aps.org/doi/10.1103/PhysRevA.83.032309}{http://link.aps.org/doi/10.1103/PhysRevA.83.032309}

%17
 \bibitem{Hoban2011}
Hoban, Matty J and Campbell, Earl T and Loukopoulos, Klearchos and Browne, Dan E,
``Non-adaptive measurement-based quantum computation and multi-party Bell inequalities''
\newblock New Journal of Physics \textbf{13}, number 2, 023014 , (2011).
\href{http://dx.doi.org/10.1088/1367-2630/13/2/023014}{http://dx.doi.org/10.1088/1367-2630/13/2/023014}.
\newblock R.~Raussendorf,
 ``Quantum computation, discreteness, and contextuality'',
arXiv:quant-ph/0907.5449, (2009).

%18
\bibitem{Mermin1990}
Mermin, N.D.
\newblock ``Quantum mysteries revisited'',
\newblock American Journal of Physics \textbf{58}, 8 731--734, (1990).
\href{http://dx.doi.org/doi:10.1119/1.16503}{http://dx.doi.org/doi:10.1119/1.16503}

%19
\bibitem{dabacon}
This was first shown to us by Dave Bacon. See e.g.~\cite{Tessier2005} for a similar construction.
%20
\bibitem{Tessier2005}
Tessier, Tracey and Caves, Carlton and Deutsch, Ivan and Eastin, Bryan and Bacon, Dave,
\newblock ``Optimal classical-communication-assisted local model of n-qubit Greenberger-Horne-Zeilinger correlations'',
\newblock Phys.~Rev.~A \textbf{72}, 032305, (2005).
\href{http://link.aps.org/doi/10.1103/PhysRevA.72.032305}{http://link.aps.org/doi/10.1103/PhysRevA.72.032305}

%21
\bibitem{Collins2004}
Collins, Daniel and Gisin, Nicolas,
\newblock ``A relevant two qubit Bell inequality inequivalent to the CHSH inequality'',
\newblock J.~Phys.~A \textbf{37}, 5, 1775--1787, (2004).
\href{http://dx.doi.org/doi:10.1088/0305-4470/37/5/021}{doi:10.1088/0305-4470/37/5/021}

%22
 \bibitem{Avis2000Revised}
D.~Avis,
``A revised implementation of the reverse search vertex enumeration algorithm''
in \emph{Polytopes - Combinatorics and Computation (Oberwolfach Seminars)} , (editors: Kalai, G. and Ziegler, G.), pp.~177--198, Birkh\"{a}user Basel, (2000). (\texttt{lrs} available at \href{http://cgm.cs.mcgill.ca/~avis/C/lrs.html}{http://cgm.cs.mcgill.ca/~avis/C/lrs.html})

%23
\bibitem{WvDMH:2010}
W.~van~Dam and M.~Howard,
``Noise Thresholds for Higher Dimensional Systems using the Discrete Wigner Function"
\newblock Phys.~Rev.~A. \textbf{83}, 032310, (2011).
\href{http://dx.doi.org/doi:10.1103/PhysRevA.83.032310}{doi:10.1103/PhysRevA.83.032310}


%24
\bibitem{Clauser1969}
Clauser, JF and Horne, MA and Shimony, A and Holt, RA
\newblock ``Proposed experiment to test local hidden-variable theories'',
\newblock Phys.~Rev.~Lett \textbf{23}, 23.880, (1969).
\href{http://dx.doi.org/doi:10.1103/PhysRevLett.23.880}{http://dx.doi.org/doi:10.1103/PhysRevLett.23.880}

%25
\bibitem{Froissart}
Froissart, M.,
\newblock ``Constructive generalization of Bell's inequalities'',
\newblock Il Nuovo Cimento B \textbf{64}, 2, 241-251, (1981).
\href{http://dx.doi.org/10.1007/BF02903286}{http://dx.doi.org/10.1007/BF02903286}

%26
\bibitem{Pitowsky2001}
Pitowsky, Itamar and Svozil, Karl,
\newblock ``Optimal tests of quantum nonlocality'',
\newblock Phys.~Rev.~A \textbf{64}, 014102, (2001).
\href{http://link.aps.org/doi/10.1103/PhysRevA.64.014102}{http://link.aps.org/doi/10.1103/PhysRevA.64.014102}

%27
\bibitem{Sliwa2003}
Sliwa, C,
\newblock ``Symmetries of the Bell correlation inequalities'',
\newblock Phys.~Lett.~A \textbf{317}, 3-4, 165--168, (2003).
\href{http://dx.doi.org/10.1016/S0375-9601(03)01115-0}{http://dx.doi.org/10.1016/S0375-9601(03)01115-0}


%28
\bibitem{Boykin2000}
Boykin, P,
``A new universal and fault-tolerant quantum basis"
\newblock Information Processing Letters \textbf{75}, 3 pp.~101--107, (2000).
\href{http://dx.doi.org/10.1016/S0020-0190(00)00084-3}{http://dx.doi.org/10.1016/S0020-0190(00)00084-3}



%29
\bibitem{Bonderson2010a}
Bonderson, Parsa and Clarke, David J.~and Nayak, Chetan and Shtengel, Kirill,
\newblock Implementing Arbitrary Phase Gates with Ising Anyons,
\newblock Phys.~Rev.~Lett. \textbf{104}, 180505, (2010).
\href{http://link.aps.org/doi/10.1103/PhysRevLett.104.180505}{http://link.aps.org/doi/10.1103/PhysRevLett.104.180505}

%30
\bibitem{Clarke2010}
Clarke, D.J.~and Shtengel, K.,
\newblock Improved phase-gate reliability in systems with neutral Ising anyons,
\newblock Phys.~Rev.~B \textbf{82}, 180519, (2010).
\href{http://link.aps.org/doi/10.1103/PhysRevB.82.180519}{http://link.aps.org/doi/10.1103/PhysRevB.82.180519}

%\bibitem{Baraban2010}
%Baraban, M. and Bonesteel, N. E. and Simon, S. H.,
%\newblock ``Resources required for topological quantum factoring'',
%\newblock Phys.~Rev.~A \textbf{81}, 062317, (2010).
%\href{http://link.aps.org/doi/10.1103/PhysRevA.81.062317}{http://link.aps.org/doi/10.1103/PhysRevA.81.062317}


%
%%12
%\bibitem{VirmaniHuelgaPlenio:2005}
%Virmani, S. and Huelga, S.F. and Plenio, M.B.,
%\newblock ``Classical simulability, entanglement breaking, and quantum computation thresholds'',
%\newblock Phys.~Rev.~A \textbf{71}, 042328, (2005).
%\href{http://link.aps.org/doi/10.1103/PhysRevA.71.042328}{http://link.aps.org/doi/10.1103/PhysRevA.71.042328}




\end{thebibliography}
\end{document}